\documentclass[conference]{IEEEtran}
\IEEEoverridecommandlockouts

\usepackage{cite}
\usepackage{amsmath,amssymb,amsfonts}
\usepackage{algorithmic}
\usepackage{graphicx}
\usepackage{textcomp}
\usepackage{pifont}
\usepackage{xcolor}
\usepackage{afterpage}
\usepackage{hyperref}
\usepackage{listings}
\usepackage{graphicx}
\usepackage{subfig}
\usepackage{hyperref}
\usepackage{soul}
\usepackage{multicol}
\usepackage{lipsum}
\usepackage{mwe}

\def\BibTeX{{\rm B\kern-.05em{\sc i\kern-.025em b}\kern-.08em
    T\kern-.1667em\lower.7ex\hbox{E}\kern-.125emX}}

\begin{document}

\title{Deep Learning-based Text-in-Image Watermarking\\
}

\author{
    \IEEEauthorblockN{
    Bishwa Karki\IEEEauthorrefmark{1}, 
    Chun-Hua Tsai\IEEEauthorrefmark{2},
    Pei-Chi Huang\IEEEauthorrefmark{1},
    Xin Zhong\IEEEauthorrefmark{1}
    }\
    \smallskip
    \IEEEauthorblockA{
    \IEEEauthorrefmark{1}
    Department of Computer Science, University of Nebraska Omaha, Omaha, NE, USA
    \\ 
    \IEEEauthorrefmark{2}
    Information Systems and Quantitative Analysis, University of Nebraska Omaha, Omaha, NE, USA
    \\ 
    \{bishwakarki, chunhuatsai, phuang, xzhong\}@unomaha.edu
    }
}

\maketitle
\thispagestyle{plain}
\begin{abstract}
In this work, we introduce a novel deep learning-based approach to text-in-image watermarking, a method that embeds and extracts textual information within images to enhance data security and integrity. 
Leveraging the capabilities of deep learning, specifically through the use of Transformer-based architectures for text processing and Vision Transformers for image feature extraction, our method sets new benchmarks in the domain. 
The proposed method represents the first application of deep learning in text-in-image watermarking that improves adaptivity, allowing the model to intelligently adjust to specific image characteristics and emerging threats. Through testing and evaluation, our method has demonstrated superior robustness compared to traditional watermarking techniques, achieving enhanced imperceptibility that ensures the watermark remains undetectable across various image contents.

\end{abstract}


\section{Introduction}
\begin{figure}[h!]
    \centering
    \vspace{-1.0em}
    \includegraphics[width=0.8\linewidth]{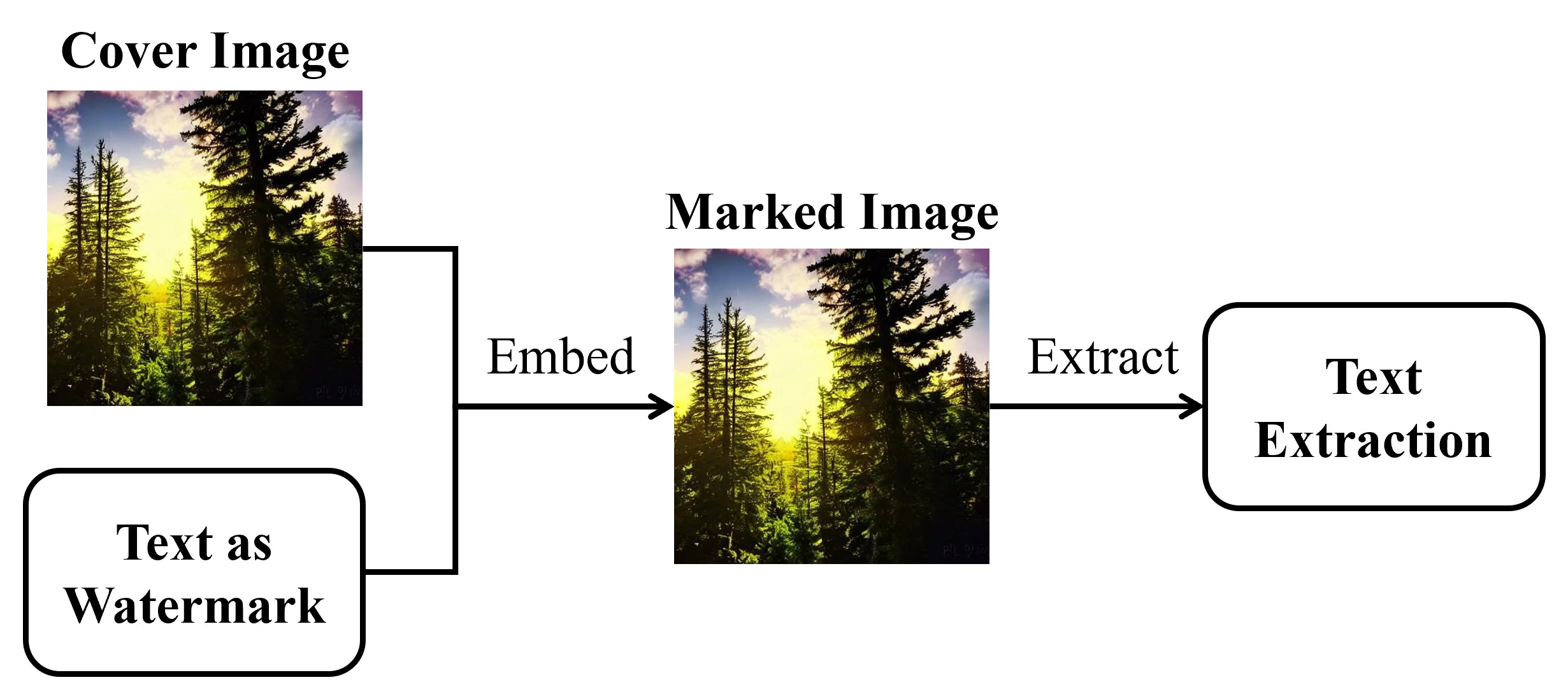}
    \vspace{-0.2em}
    \caption{Text-in-Image Watermarking.}
    \vspace{-1.0em}
    \label{fig:general}
\end{figure}

The concept of watermarking within digital media pertains to the technique of imperceptibly embedding information into a host medium to assert copyright, authenticate, or convey additional data without compromising the medium's original quality or appearance. 
Text-in-image watermarking, a specific subset of this domain, involves the insertion of textual information into digital images in such a manner that it remains invisible to human vision. 
This process is designed to ensure that the integrity and aesthetics of the cover image are maintained, while still embedding a watermark that can be retrieved or detected through specific, authorized methods (see Fig.~\ref{fig:general}).

\noindent The primary objective of text-in-image watermarking is to achieve a balance between imperceptibility and robustness~\cite{ref_1, hidden, ref_3}; the watermark should not only be undetectable during casual observation but also robust against various forms of digital manipulation and distortion. 

\noindent \textbf{Deep Learning-based Watermarking.}
As the field of image watermarking seeks to improve both its robustness and imperceptibility, deep learning has proven to be an important method~\cite{zhong2023brief}.   
By leveraging the sophisticated pattern recognition and learning capabilities inherent in deep learning models, watermarking algorithms are endowed with unprecedented adaptability. 
These models excel in discerning the intricate features of cover images, allowing for the embedding of watermarks in a manner that is both subtle and robust to attacks. 
Furthermore, the dynamic nature of deep learning approaches facilitates a continuous evolution in response to new challenges and threats, heralding a significant leap forward in the field of digital watermarking. 
This shift towards deep learning-based solutions not only marks a milestone in achieving a delicate balance between imperceptibility and robustness but also redefines the boundaries of what is achievable in protecting digital content against unauthorized use and manipulation.

\noindent \textbf{Text as Watermark.} 
The integration of text within images presents unique advantages and challenges. 
Text watermarks excel in bolstering the security of both the embedded textual information and the cover image itself. This dual-part protection is paramount in scenarios requiring the authentication of the document's origin while safeguarding the embedded message. 
Such a mechanism is particularly crucial in legal documentation, intellectual property rights, and confidential communications, where verifying the authenticity of the document and its content is as important as protecting the information itself. 
Furthermore, in digital media distribution, this approach aids in combating unauthorized reproduction and manipulation, ensuring that the creator's rights are preserved and the content remains tamper-proof.

Despite the proliferation of research in separate domains of cover-image~\cite{zhong2023brief} and cover-text watermarking~\cite{ou2003text, rizzo2019fine, kamaruddin2018review, anitw}, the amalgamation of text within images as a unified watermarking approach poses significant challenges. One such challenge arises from the inherent multi-modality of combining textual and visual information, where maintaining the imperceptibility and integrity of both the text and the image necessitates sophisticated embedding techniques. 
The complexity is further compounded by the need to ensure that the watermark remains robust against various forms of digital manipulation without detracting from the visual quality of the cover image. 
Consequently, advancing text-in-image watermarking schemes that adeptly navigate these multimodal intricacies is of great interest.

Traditional text-in-image watermarking methods~\cite{ref_5} encounter several critical drawbacks that impede their effectiveness in advanced digital media applications. 
Firstly, the absence of learning integration results in less adaptivity to the varying nature of image content and emerging threats, rendering these methods less flexible. 
Secondly, such conventional approaches exhibit lower robustness, struggling to maintain watermark integrity against sophisticated forms of manipulation and compression. 
Thirdly, it is hard for these methods to achieve high imperceptibility, especially in complex or detailed image scenarios, where the watermark might become perceptible and affect the visual quality. 
These limitations underscore the necessity for innovative approaches that can overcome these challenges.

To this end, we introduce a pioneering text-in-image watermarking method leveraging deep learning, marking a significant advancement in the domain. 
Our contributions are manifold: 
(1) This method represents the first application of deep learning in text-in-image watermarking, significantly enhancing adaptivity to intelligently adjust to the specific characteristics of each image and evolving digital threats. 
(2) It exhibits robustness that has been rigorously tested and systematically evaluated, demonstrating superiority compared to traditional methods. 
(3) Our approach achieves better imperceptibility, ensuring the watermark remains undetectable in various image contents, thus preserving the pristine quality of the original image. 
 
The structure of the paper is organized as follows: Section~\hyperref[sec:related]{II} discusses related works in the field. Section~\hyperref[sec:proposed]{III} details the proposed watermarking method. Data, experimental setups, and their analyses are presented in Section~\hyperref[sec:experiments]{IV}. Finally, Section~\hyperref[sec:conclusion]{V} discusses the conclusions drawn from our research.
 
\section{Related Works}
\label{sec:related}
This section provides a succinct review of the relevant literature, with the discussion bifurcated into an overview of deep learning applications in image watermarking, and an examination of traditional text-in-image watermarking techniques.

\subsection{Image Watermarking and Deep Learning}
Deep learning has revolutionized the field of image watermarking by introducing models that are inherently more adaptive and robust to a variety of attacks~\cite{zhong2023brief}. 
The emergence of deep learning as powerful tools for feature extraction and pattern recognition has led to the development of sophisticated watermarking schemes. These schemes are typically characterized by an embedder-extractor framework, where the embedder network is trained to insert a watermark in such a way that it is imperceptible to the human eye, yet can be accurately extracted by the corresponding extractor network, even in the presence of noise~\cite{hidden, ref_1, ref_3, redmark, mbrs}. 

Such neural network-based approaches benefit from the ability to learn complex representations of image data, which traditional algorithms cannot easily capture. This allows for watermarks that are intricately interwoven with the content of the image, enhancing security and robustness. 
Moreover, these models can be trained end-to-end, allowing for simultaneous optimization of both embedding and extraction processes. The adaptability of deep learning models to different content types and distortions has established a new standard in the watermarking domain, particularly for applications requiring high levels of security and discretion.

\vspace{-0.5em}
\subsection{Text-in-Image Watermarking}
Traditional methods have laid the groundwork for text-in-image watermarking by integrating techniques such as Discrete Cosine Transform (DCT) and Discrete Wavelet Transform (DWT) for embedding text securely within images. 
Gupta and Khunteta~\cite{ref_5} tackled the challenge of text-in-image watermarking by proposing an algorithm that leverages DCT and DWT. Their method focuses on the text watermark's security and integrity. 
Singh~\cite{singh2017improved} develops a robust hybrid multiple watermarking technique for medical images and multimedia objects, employing a fusion of DWT, DCT, and Singular Value Decomposition (SVD) to embed multiple watermarks simultaneously for identity authentication and added security. This approach involves an embedding process that enhances the imperceptibility and robustness of the watermark, including the use of encryption for text watermarks considered as Electronic Patient Record data.
Anand and Singh~\cite{anand2020improved} introduced a watermarking technique designed to secure patient records transmitted over networks, crucial for tele-health services. Their method embeds multi-watermarks in medical images using the DWT-SVD domain, enhanced with Hamming code for text watermarks, and employs a combination of Chaotic-LZW for encryption and compression. 

While established text watermarking techniques have contributed significantly to the development of digital watermarking, they present notable limitations in adaptivity, robustness, and imperceptibility. 
Our proposed method, merging these considerations, introduces deep learning to text-in-image watermarking for the first time. 
This innovative approach not only acknowledges the foundational importance of traditional methods but also significantly advances the field by addressing these limitations and setting new benchmarks for watermarking performance, representing a major shift towards novel watermarking solutions.

\section{The Proposed Method}
\label{sec:proposed}
\begin{figure*}[!htb]
    \centering
    \vspace{-1.5em}
    \includegraphics[width=0.95\linewidth]{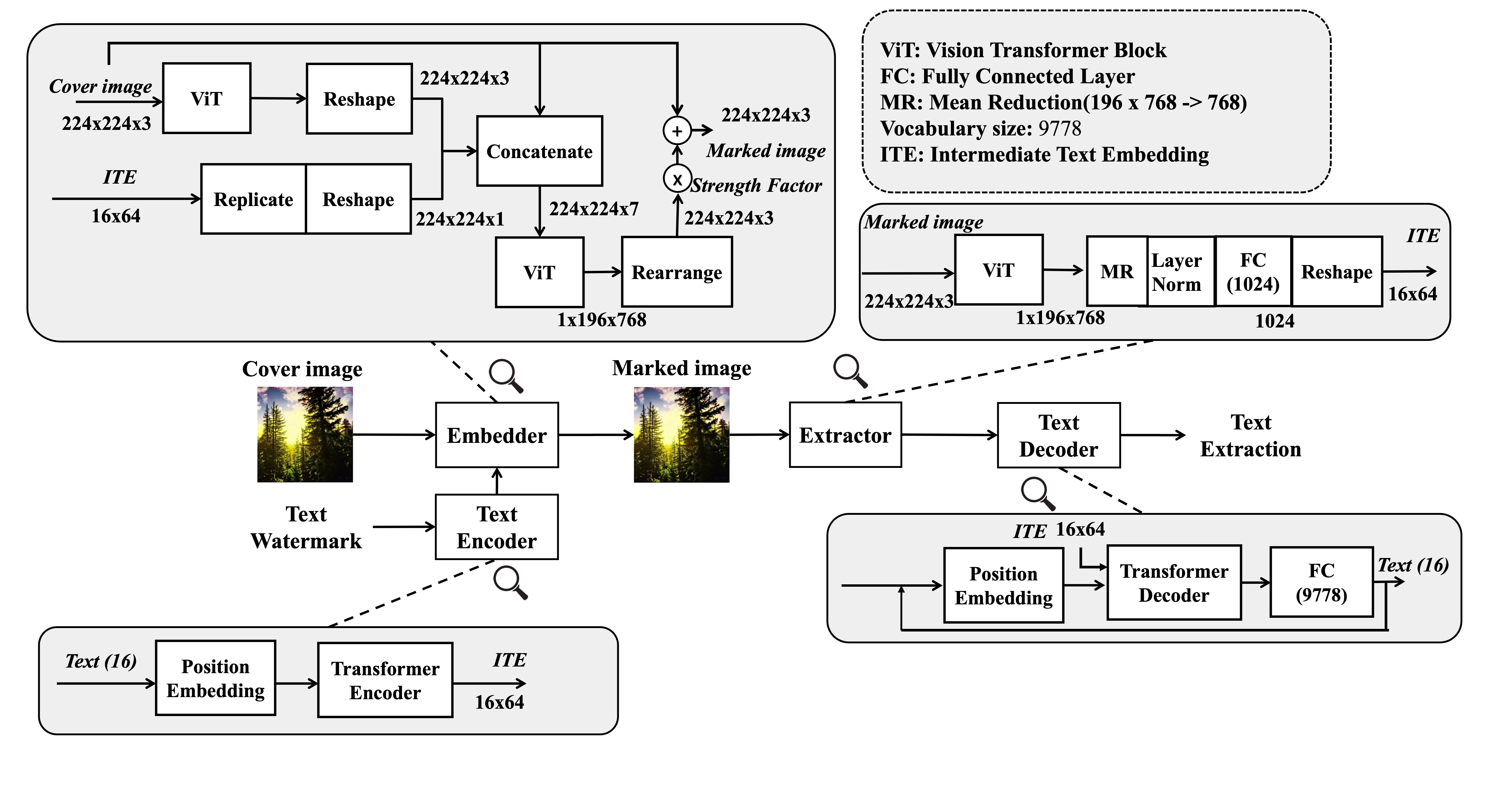}
    \vspace{-2.0em}
    \caption{Overview of the proposed method}
    \label{fig:Architecture}
    \vspace{-1.5em}
\end{figure*}

This section elaborates on the architecture of the proposed model, alongside its loss and training methodology. 
Although the model is eventually trained as a cohesive deep learning network, for clarity in discussion, it is conceptually segmented into four major components: the encoder, decoder, embedder, and extractor. Fig.~\ref{fig:Architecture} depicts the overall architecture and provides an overview of our text-in-image watermarking scheme.

\vspace{-0.5em}
\subsection{The Encoder and Decoder Networks}
The widespread adoption of Transformer-based architectures, known for their exceptional performance in linguistic tasks like summarization and machine translation, motivates our use of Transformer design in our encoder-decoder network.

The encoder and decoder constitute a Transformer~\cite{transformer} based auto-encoder, in which the encoder's objective is to map every word in a sentence $\mathbf{W} = (x_1, \ldots, x_n)$ to a contextual representation $\mathbf{Z} = (z_1, \ldots, z_n)$ and the decoder's goal is to use the representation $\mathbf{Z}$ that the encoder passes along to re-generate the original input sentence $\mathbf{W}$, training on the next word prediction objective~\cite{devlin2018bert}. The text encoder uses positional encoding to create a 64-dimensional embedding after receiving a token representation of a sentence with a fixed length of 16 words, as illustrated in Fig. \ref{fig:Architecture}. On the other hand, the Transformer's feed forward layer uses a dimension of 512. The words in the input sentence are represented in $16 \times 64$ dimensions by the encoder, which consists of three layers.

Text decoder as illustrated in Fig.~\ref{fig:Architecture} receives an embedding of $16 \times 64$, processes it via three layers of decoder blocks, and adds a fully connected layer on top of the decoder to predict vocabulary in an auto-regressive fashion to produce output sentence $\mathbf{W^{\prime}} = (x_1^{\prime}, \ldots, x_n^{\prime})$ similar to $\mathbf{W}$. This process is similar to that of the encoder. In the Transformer's feed forward layer, an embedding size of 512 is utilized.  

\vspace{-0.5em}
\subsection{The Embedder and Extractor Networks}
Previous methods have predominantly used convolutional layers for deep learning-based watermarking \cite{hidden, redmark, ref_1}. Inspired by the state-of-the-art achievements of ViT \cite{vit} in various vision tasks, our work uses ViT as the feature extractor in both the embedder and extractor networks, with modifications to the original architecture by omitting the CLS token and classification head.

The embedder network is responsible to embed the text representation $\mathbf{Z}$ into the cover image $\mathbf{I_{c}}$ and make it imperceptible while the extractor takes in the marked image $\mathbf{I_{m}}$ produced by the embedder and produce back the textual representation $\mathbf{Z^{\prime}}$. The extractor must be robust enough to handle distortions. 

The embedder receives a $16 \times 64$ representation of the text  $\mathbf{Z}$  and a $224 \times 224 \times 3$ cover image  $\mathbf{I_{c}}$ . First, we use ViT, an embedding dimension of 768 and three layers with patch size of 16, to learn the image features with three channels. Then, we reshape the feature back to its original size of $224 \times 224 \times 3$. In the embedder, the text feature is also repeated and reshaped to create a vector with dimensions of $224 \times 224 \times 1$. And we create a representation of the $224 \times 224 \times 7$ form by concatenating those features in a channelwise manner with the cover image. This feature set is further processed using ViT, which has three layers and a 786 embedding dimension. However, in order to learn the representation of the embedded picture and text features simultaneously, it now requires a channel of seven instead of three as it did previously. ViT serves as a feature embedder here as well. Lastly, we reshaped the learnt characteristics to the original shape $224 \times 224 \times 3$ and add to the cover image $\mathbf{I_{c}}$ by multiplying $\mathbf{I_{m}}$ using a strength factor forming  marked image $\mathbf{I_{m}}$.

The representation of the text encoded in the cover image is extracted using the extractor network. The extractor receives a marked image $\mathbf{I_{m}}$ of shape $224 \times 224 \times 3$, as seen in Fig.~\ref{fig:Architecture}, and sends it to ViT, which consists of three layers and an embedding size of 768, to generate a learnt representation of shape $1 \times 196 \times 768$ using a patch size of 16 in ViT and using mean reduction, $1 \times196 \times 768$ is reduced to a shape of $1 \times 768$. A Fully Connected Layer processes this feature to transform a dimension of 768 into 1024. The final output is reconfigured to appear like a $16 \times 64$ text representation $\mathbf{Z^{\prime}}$.

\vspace{-0.75em}
\subsection{Training and Loss Functions}
\vspace{-0.5em}
During training, we aim to balance text fidelity and image quality by optimizing a set of loss functions. 
A two-phase training strategy~\cite{hidden, doc_img} is employed, addressing the issue of training imbalance that typically arises when initiating training from scratch. This approach allows for a more controlled and effective optimization of the network's performance on both fronts.

\subsubsection{Encoder-Decoder Networks Pre-training}
\begin{figure}[h!]
    \centering
    \vspace{-0.75em}
    \includegraphics[width=0.8\linewidth]{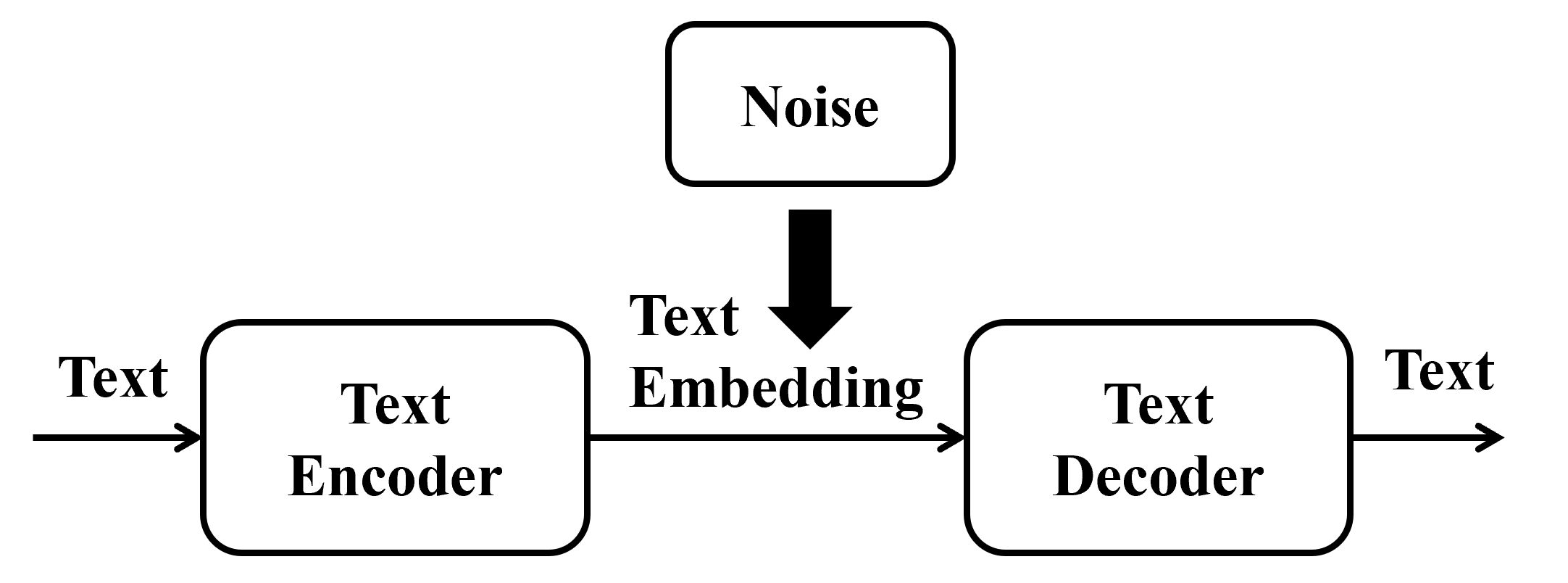}
    \vspace{-0.2em}
    \caption{Text Encoder-Decoder Pre-training}
    \vspace{-1.0em}
    \label{fig:enc_dec}
\end{figure}

Initially, our approach focuses on pre-training the encoder-decoder model (see Fig.~\ref{fig:enc_dec}) for precise input sentence regeneration, establishing a robust foundation for the model.  
We adopt an auto-regressive technique for the initial training of the text encoder and decoder, employing binary cross-entropy loss: 
\vspace{-0.5em}
\begin{equation}
    L_1 = - \sum_{i=1}^{C} x_i \log(x_i^{\prime}),   
    \label{eqn:entropy}
\end{equation}
where $x_i$ and $x_i^{\prime}$ are representations of actual word and probability of predicted word respectively, while $C$ denotes total number of unique words in the vocabulary.

We have introduced various noises to the text embedding (detailed in Section~\ref{sec: Ablation Study pretraining}). Incorporating noise during pre-training is pivotal, markedly enhancing the model's robustness. This strategy compels the text decoder to learn to distinguish and rectify noisy representations, effectively training it to grasp the intrinsic structure of the text embedding with noise. Such a capacity to discern and correct distortions ensures the model's adeptness at dealing with diverse perturbations, significantly fortifying its robustness. 

Unlike previous approaches that primarily focus on adding noise to the marked images~\cite{ref_1, hidden} to test robustness, our method emphasizes adding noise directly to the text embedding. This innovation not only challenges the decoder in novel ways but also significantly increases the generalizability of the model's robustness by ensuring it is not merely image perturbations it can withstand, but also variations at the feature level of the text watermark itself.

\subsubsection{Training the entire Network}
\label{sec:training}
Following this encoder-decoder pre-training phase, we proceed to train the entire network, with a focus on the embedder and extractor components, while learning the encoder and decoder with a small learning rate. This method ensures a gradual, focused enhancement of the network's performance, leveraging the foundational strengths established during the initial pre-training and entire training stage.

The total training loss of our comprehensive network, which includes an encoder, embedder, extractor, and decoder components, is defined by equation~\ref{eqn:total_loss}, incorporating a blend of four distinct terms. This loss function is designed to optimize the network across various aspects of performance:
\vspace{-0.5em}
\begin{equation}
    L_{total} = \lambda_1 L_1 + \lambda_2 L_2 + \lambda_3 L_3 - \lambda_4 \cdot \text{SSIM}.
    \label{eqn:total_loss}
\end{equation}
Here, \(L_1\) assesses the precision of the text content regenerated by the decoder. The terms \(L_2\) and \(L_3\) correspond to the Mean Square Error (MSE), where \(L_2\) calculates MSE between the cover image $\mathbf{I_{c}}$ and the marked image $\mathbf{I_{m}}$, and \(L_3\) measures the discrepancy between the text embedding $\mathbf{Z}$ produced by the encoder and the embedding $\mathbf{Z^{\prime}}$ extracted by the extractor, as expressed in the following equations:
\vspace{-0.5em}
\begin{equation}
    L_2 = \frac{1}{n} \sum_{i=1}^{n} (I_{c} - I_{m})^2,
    \label{eqn:mse_image}
\end{equation}

\vspace{-0.5em}
\begin{equation}
    L_3 = \frac{1}{n} \sum_{i=1}^{n} (Z - Z^{\prime})^2. 
    \label{eqn:mse_emb}
\end{equation}
Additionally, Structural Similarity Index (SSIM) between the cover and marked images is integrated into the loss function to account for the perceptual quality of the watermarked images:
\begin{equation}
    SSIM(I_{c}, I_{m}) = \frac{(2\mu_c\mu_m + C_1)(2\sigma_{cm} + C_2)}{(\mu_c^2 + \mu_m^2 + C_1)(\sigma_c^2 + \sigma_m^2 + C_2)},
    \label{eqn:ssim}
\end{equation}
where \(\mu_c\) and \(\mu_m\) represent the mean intensities of $\mathbf{I_{c}}$ and $\mathbf{I_{m}}$ respectively; \(\sigma_c^2\) and \(\sigma_m^2\) denote their variances, while \(\sigma_{cm}\) is the covariance between the two images. Constants \(C_1\) and \(C_2\) are introduced to stabilize the division with small denominators, defined as \(C_1 = (k_1L)^2\) and \(C_2 = (k_2L)^2\), where \(L=255\), \(k_1=0.01\), and \(k_2=0.03\) are the default values.

\section{Experiments and Analysis}
\label{sec:experiments}
This section presents the experimental evaluation of our proposed method, offering a detailed analysis from various perspectives. 
We begin with the Dataset and Implementation Details, providing the necessary background for understanding our experimental framework. 
Next, we discuss the Training and Testing Results to evaluate the model's performance on accuracy and efficiency metrics. 
The Robustness section assesses our method's robustness to different types of perturbations, emphasizing its reliability under various conditions. 
A comparison with related approaches illustrates our model's competitive edge. Finally, an ablation study highlights the contribution of individual components to the overall performance. 

\subsection{Dataset and Implementation Details}
For the training and evaluation of our proposed model, we leveraged the Microsoft COCO Dataset~\cite{lin2014microsoft} and the Multi30K Dataset~\cite{elliott2016multi30k}, focusing on their rich visual and textual content, respectively. 
The Microsoft COCO Dataset, known for its diversity in image content, provided a set of 19,959 images for training and 1,000 images for validation, all resized to $224 \times 224$ pixels and normalized to a range between 0 and 1. 
From the Multi30K Dataset, we extracted English sentences, aligning the quantity of textual data with the image dataset. Sentences were truncated or padded to ensure a uniform length of 16 words.

The proposed model is trained on a corpus of 19,959 cover images paired randomly to a sentence. The images were rescaled to a uniform dimension of $224 \times 224 \times 3$, while sentences were adjusted to a maximum of 16 words for compatibility with our model's input requirements.
We found that embedding up to 16-word segments, adjusting for redundancy and embedding dimensions, was most effective. By repeating text embedding 49 times with 64 dimensions, we improved our model's performance and robustness.

We employed the Adam optimizer for both the encoder and decoder, setting the learning rate to \(1e-4\) during pre-training and \(1e-8\) for entire network training phases. For other components of the network, a learning rate of \(1e-4\) was maintained. Additionally, during the pre-training stage, various noise was introduced to the text embeddings to enhance model robustness and a strength factor 0f 0.8 is being used.

\vspace{-0.5em}
\subsection{Training \& Testing Results}
\begin{figure}[h!]
    \centering
    \vspace{-1.0em}
    \includegraphics[width=0.5\linewidth]{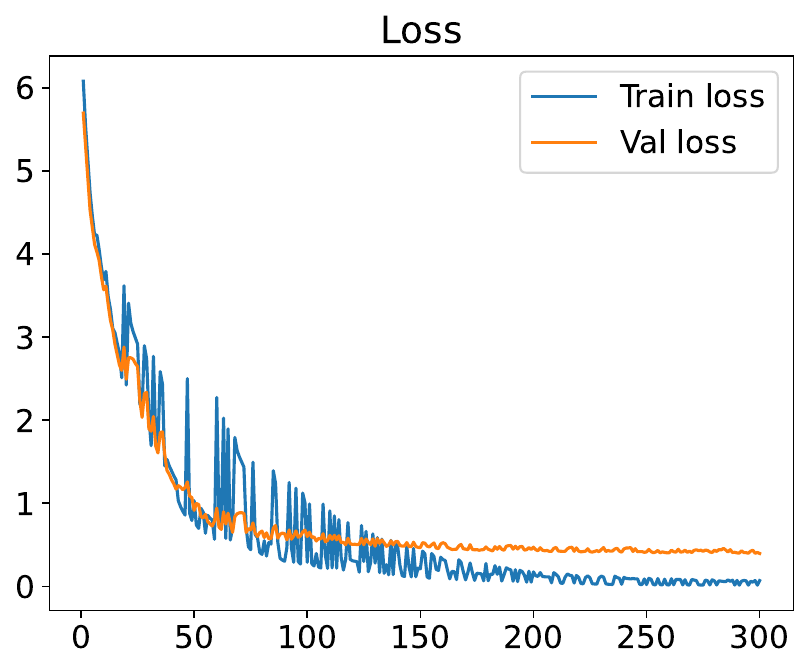}
    \vspace{-0.2em}
    \caption{Pretraining loss}
    \vspace{-1.0em}
    \label{fig:pretraining_loss}
\end{figure}

\begin{figure}[h!]
\vspace{-0.5em}
\centering
\begin{tabular}{cc}
 \includegraphics[width=.22\textwidth]{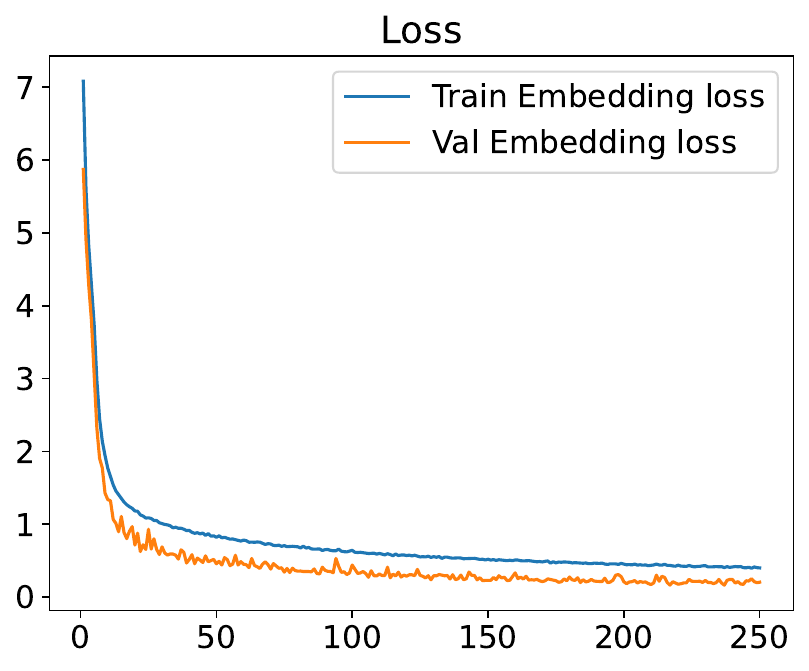}&
 \includegraphics[width=.22\textwidth]{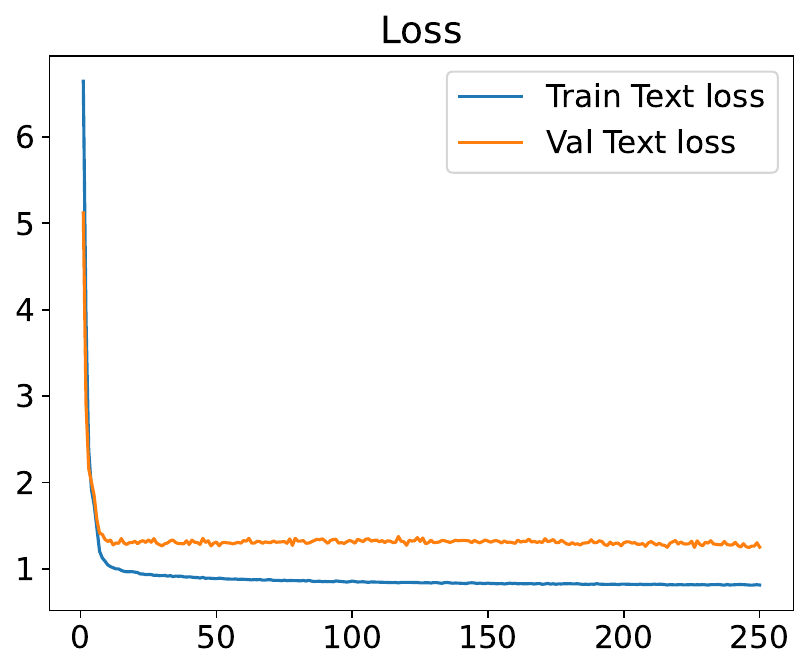}
\end{tabular}
\begin{tabular}{cc}
 \includegraphics[width=.22\textwidth]{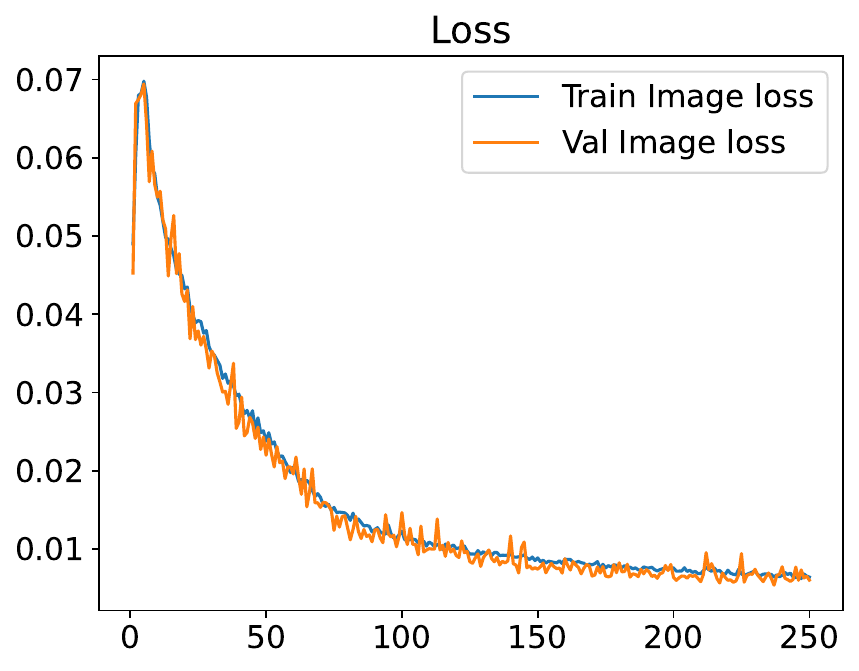}&
 \includegraphics[width=.22\textwidth]{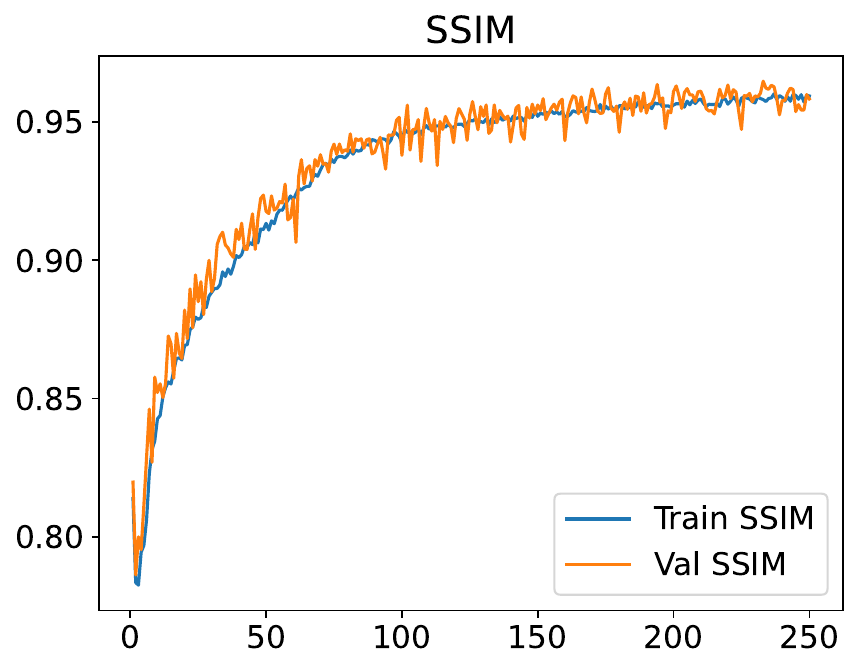}
\end{tabular}
\vspace{-0.5em}
\caption{Entire network training loss} 
\vspace{-1.0em}
\label{fig:training_loss}
\end{figure}

We followed our proposed scheme as outlined in Section~\ref{sec:proposed} and training details in Section~\ref{sec:training} for training. 
The training and validation loss of encoder-decoder pre-training and entire network training is depicted in Fig.~\ref{fig:pretraining_loss} and Fig.~\ref{fig:training_loss} respectively and has demonstrated a well-fitting model. Fig.~\ref{fig:training_loss} shows the progression of Embedding loss as defined in equation~\ref{eqn:mse_emb} (top-left), Text loss as defined in equation~\ref{eqn:entropy} (top-right), MSE (equation~\ref{eqn:mse_image}) (bottom-left) and SSIM (equation~\ref{eqn:mse_emb}) (bottom-right) throughout the training process.  After pre-training for 300 epochs, the encoder-decoder model achieved a Bilingual Evaluation Understudy (BLEU) score of 94.83\% in validation data. When using this pre-trained model in training entire network achieved a BLEU score of 91.39\% and SSIM of 97.95\% on validation data.

\begin{figure}
    \centering
    \vspace{-1.0em}
    \includegraphics[width=0.95\linewidth]{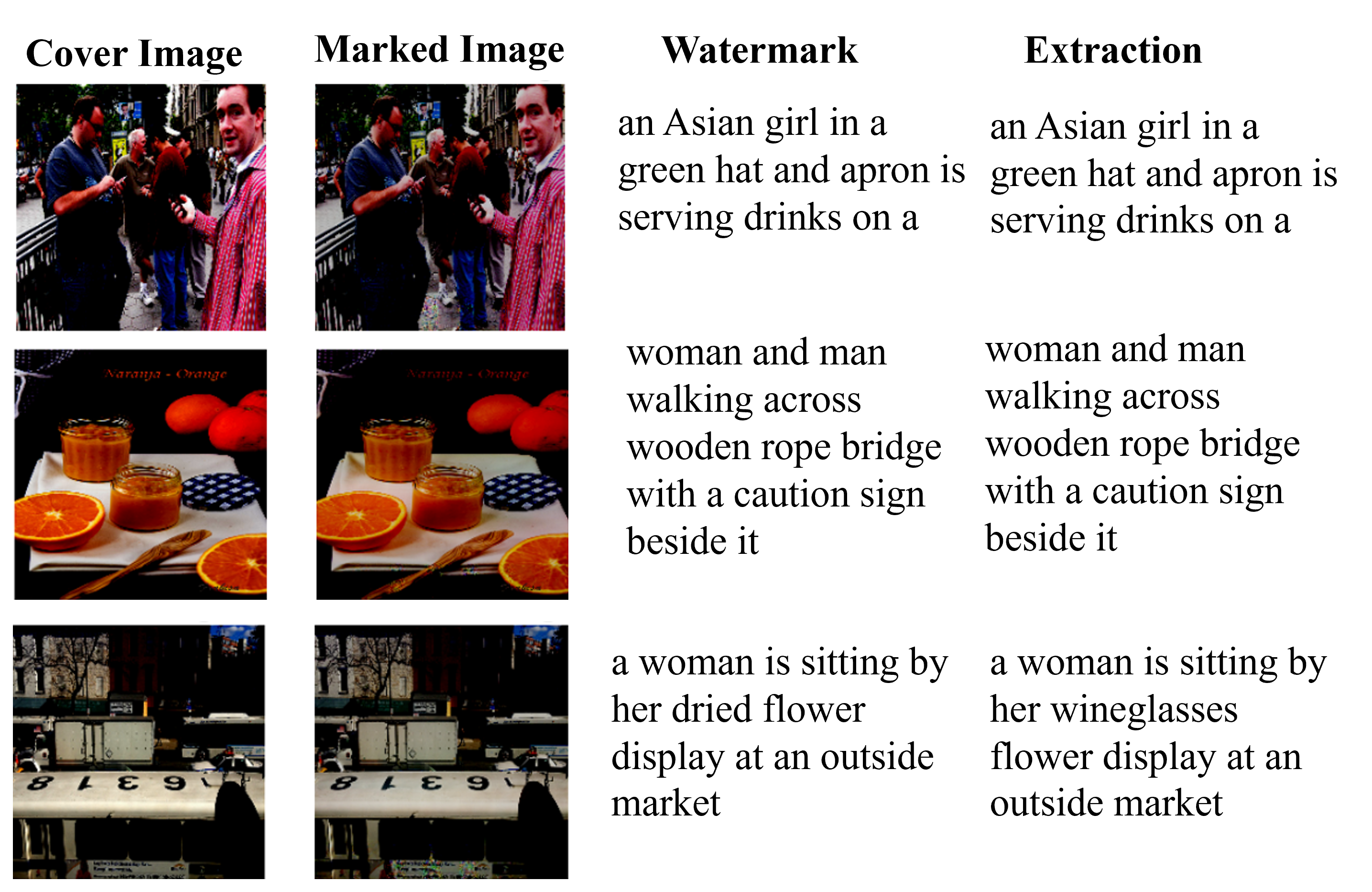}
    \vspace{-0.5em}
    \caption{Examples of cover image (1st column), marked image (2nd column), watermark text (3rd column), and extraction (4th column)}
    \vspace{-1.5em}
    \label{fig:testing_example}
\end{figure}

The model notably achieved a testing SSIM of 97.33\% and BLEU score of 91.13\% on 1000 testing samples. 
Fig.~\ref{fig:testing_example} shows a few examples of cover and marked images along with arbitrary text watermarks with the extraction. 

The balance between image quality and text clarity is managed through the weighting in the loss function, as detailed in equation~\ref{eqn:total_loss}. Prioritizing the MSE of the image, as defined in equation~\ref{eqn:mse_image}, enhances image quality, whereas emphasizing the text loss, per equation~\ref{eqn:entropy}, improves text clarity. 
We assign weights of 0.8, 3, 0.5, and 0.01 to $\lambda_1$, $\lambda_2$, $\lambda_3$, and $\lambda_4$ respectively in the equation~\ref{eqn:total_loss}.

\subsection{Robustness}

\begin{figure}[!ht]
    \centering
    \vspace{-1.0em}
    \includegraphics[width=0.75\linewidth]{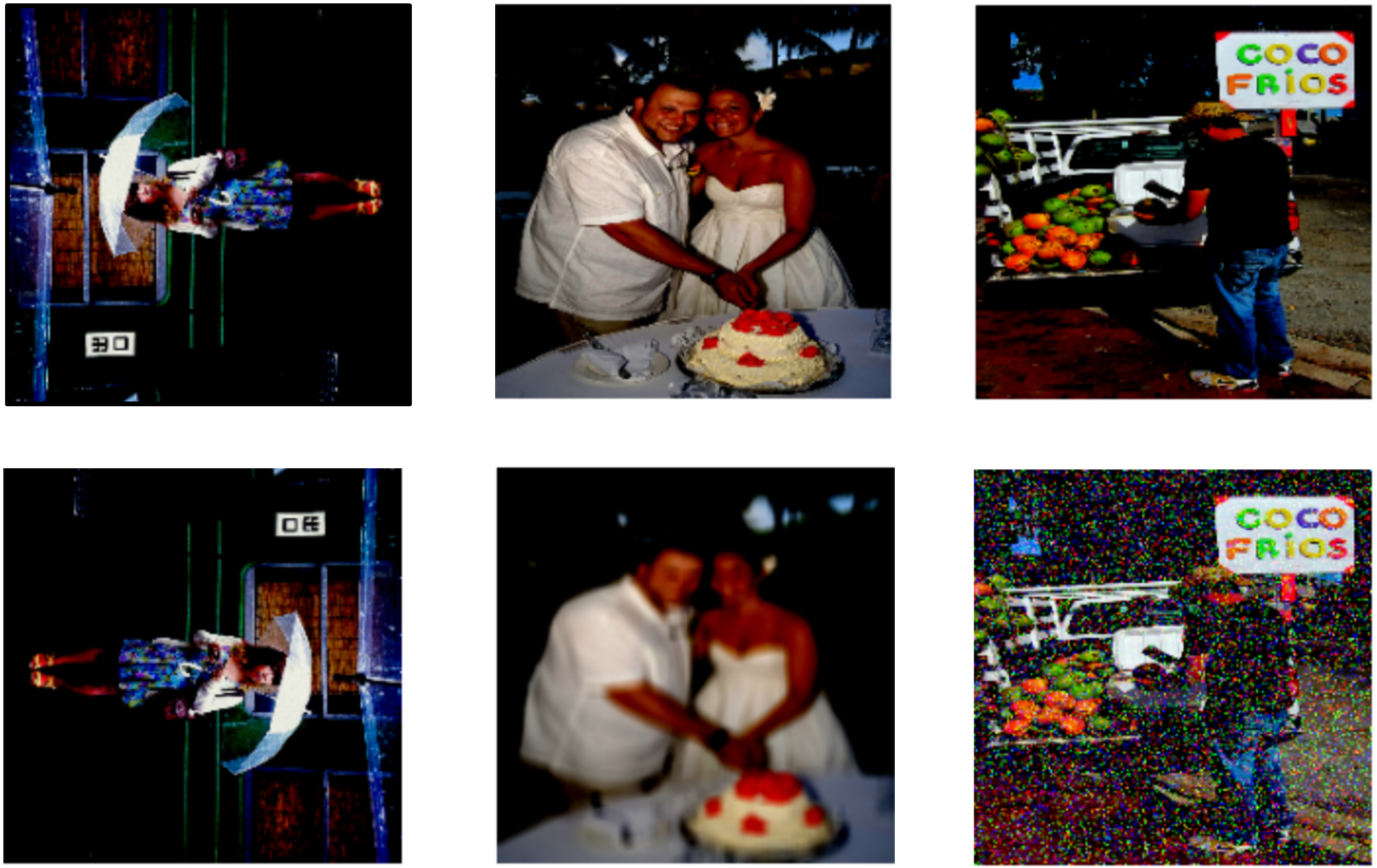}
    \vspace{-0.5em}
    \caption{Several illustrative instances of the marked images (top row) and their distortion (bottom row). The techniques are applied from left to right: Rotation, Gaussian Blur, Salt \& Pepper, respectively.}
    \vspace{-1.0em}
    \label{fig:noise_example}
\end{figure}

The robustness of our proposed text-in-image watermarking method is evaluated using a validation set of 1000 images and sentences. The method shows enhanced robustness, successfully extracting watermarks from images subjected to severe distortions, thus affirming its performance under challenging conditions. 
Examples of marked images and the applied distortions are presented in Fig.~\ref{fig:noise_example}. 

\begin{figure}[h!]
\vspace{-0.5em}
\centering
\begin{tabular}{cc}
 \includegraphics[width=.22\textwidth]{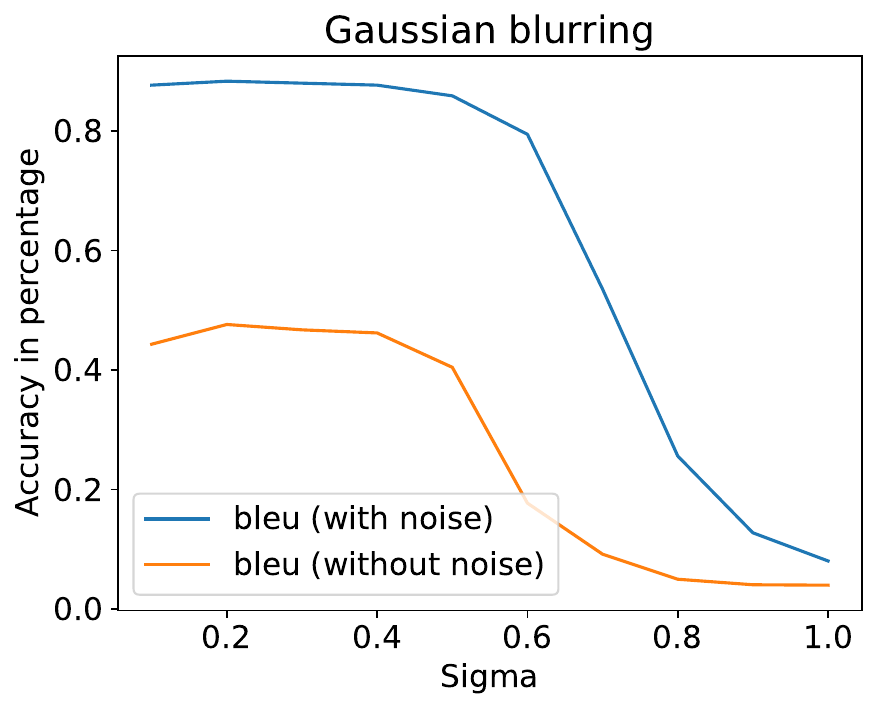}&
 \includegraphics[width=.22\textwidth]{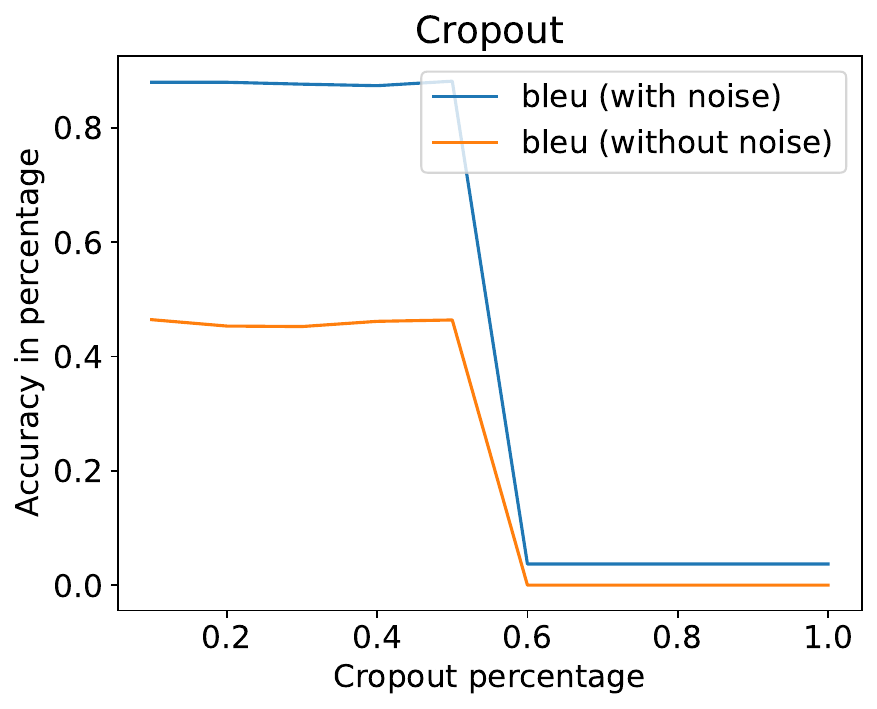}
\end{tabular}
\begin{tabular}{cc}
 \includegraphics[width=.22\textwidth]{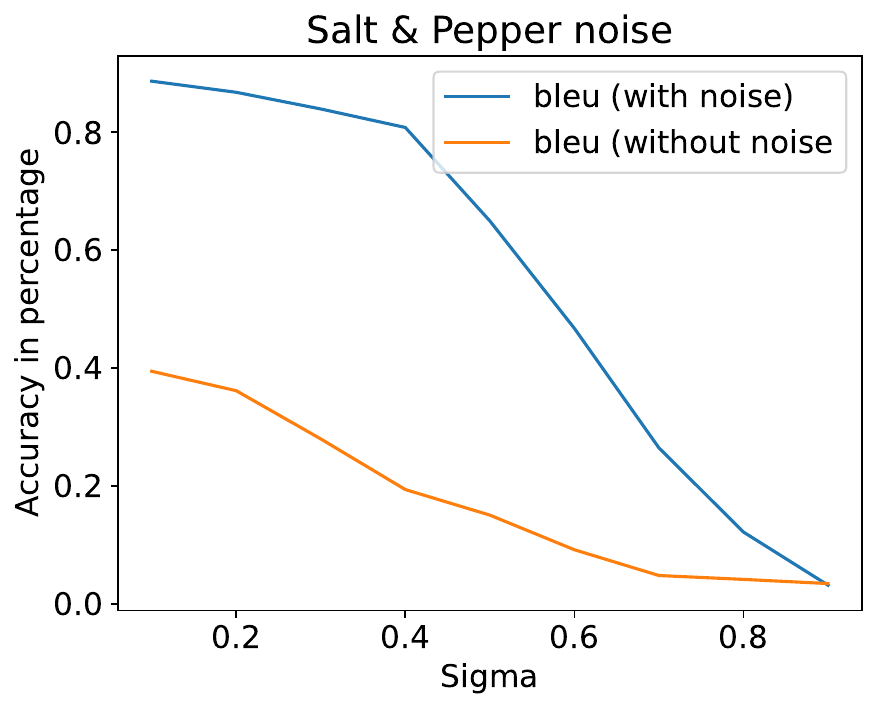}&
 \includegraphics[width=.22\textwidth]{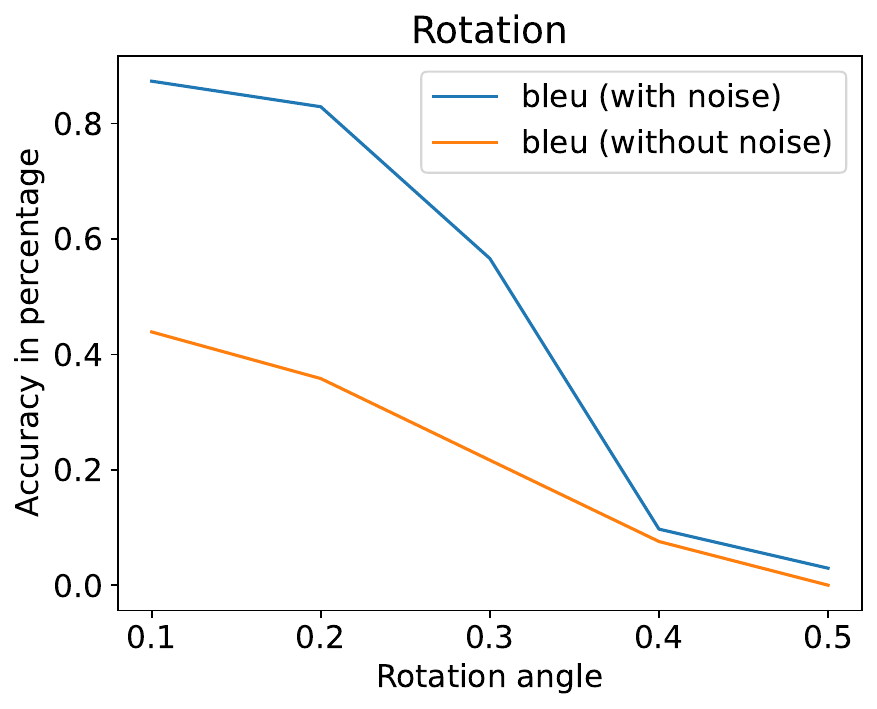}
\end{tabular}
\begin{tabular}{cc}
 \includegraphics[width=.22\textwidth]{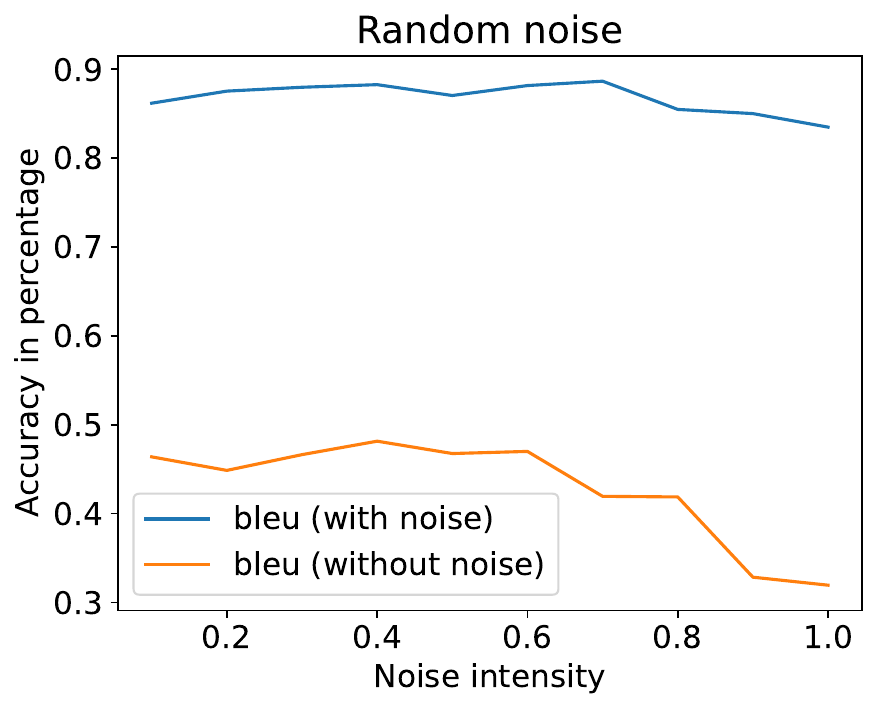}&
\end{tabular}
\vspace{-0.5em}
\caption{Noise testing}
\vspace{-1.5em}
\label{fig:noisy_pretraining_robustness}
\end{figure}

Fig.~\ref{fig:noisy_pretraining_robustness} illustrates the method's performance against increasing distortion levels, maintaining high BLEU scores across various augmentations. Further, the method demonstrated significant tolerance to escalating noise levels, with performance decreasing only slightly with intensifying noise. This performance underscores the proposed method's ability to preserve watermark integrity across a spectrum of distortions.

\subsection{Comparative Analysis}

In this study, we compare our proposed deep learning-based method against the model presented in~\cite{ref_5}, selected for its relevance and prominence in the current literature on text-in-image watermarking. The absence of deep learning approaches in this niche necessitates our comparison with traditional methods, making~\cite{ref_5} the most suitable benchmark. This choice is predicated on the reason that~\cite{ref_5} outlines a general text-in-image watermarking technique not limited to any specific domain, thus providing a broad basis. 

Alongside the method in~\cite{ref_5}, we evaluated the performance of a Gated Recurrent Units (GRU) based architecture for text encoding and decoding within our framework. This variant replaces the Transformer block in Fig.~\ref{fig:Architecture} with GRUs to assess the effectiveness of different neural models in our watermarking scheme. This comparison aims to highlight the advantages of employing Transformer blocks for enhancing text-in-image watermarking performance. 

The outcomes, detailed in Table~\ref{tab:compare}, present MSE, SSIM, and BLEU scores, underscoring our proposed method's superiority. Our approach notably excels in maintaining the cover image's quality and ensuring high accuracy in text extraction.

\begin{table}[ht]
\vspace{-0.5em}
\centering
\caption{A quantitative comparison between the proposed scheme and Gupta's scheme.}
\begin{tabular}{|c|c|c|c|}
\hline
Model & MSE & SSIM & BLEU \\
\hline
Gupta \cite{ref_5} & 0.26578 & 0.45578 & N/A \\
GRU-based & 0.012 & 0.85 & 0.71\\
Proposed method & 0.0020 & 0.9733 & 0.9113\\
\hline
\multicolumn{4}{l}{(Note:  N/A denotes "Not Applicable" that the metric was not covered.)}
\end{tabular}
\label{tab:compare}
\vspace{-0.5em}
\end{table}

\vspace{-0.5em}
\subsection{Ablation Study}
\label{sec: Ablation Study pretraining}
This experiment highlights the advantages of incorporating noise during the pretraining phase of the encoder and decoder, supporting our hypothesis that noise addition enhances the decoder's robustness, enabling it to better generalize and handle noise when integrated into the full watermarking network. Fig.~\ref{fig:enc_dec} illustrates the pretraining process, including noise addition to the intermediate text embeddings within the encoder-decoder framework. 
Introducing noise types during pretraining that may mimic potential disturbances in marked images has led to increased robustness across the entire network. Specifically, pretraining the encoder-decoder model with noise has improved the performance significantly, achieving an SSIM of 97.33\%, a Peak Signal-to-Noise Ratio (PSNR) of 33.16dB, and a BLEU score of 91.13\% upon integration with the complete network. 
In contrast, pretraining without noise introduction resulted in lower performance metrics: an SSIM of 96.81\%, a PSNR of 32.41dB, and a BLEU score of 89.13\%.

During the pretraining of the encoder-decoder, noise was specifically added to the text embeddings to enhance robustness, a strategy that generalized well to different scenarios. 
Even when distinct noises were applied to the marked images during testing, the decoder was able to effectively extract the watermark. 
To facilitate this generalization from the text embeddings to manage distortions introduced by the marked image, we engineer the noise to the encoder-decoder pretraining by replacing 30\% of the text embedding with zeros, applying 1D Gaussian blurring to the text embedding vector, and introducing random Gaussian noise (mean = 0, standard deviation = 2). 
During pretraining, we diversified the robustness training by randomly selecting a type of noise from the aforementioned methods and adding it to the intermediate text embedding vector in each iteration (Fig.~\ref{fig:enc_dec}). 
This methodological inclusion of noise is based on the understanding that exposing the model to a variety of perturbations enhances its ability to generalize from the specific conditions encountered during training to a broader range of scenarios. Such a strategy leverages the concept that robustness against one type of distortion can improve robustness against others, thereby broadening the model's capabilities. 
As evidenced in Fig.~\ref{fig:noisy_pretraining_robustness}, models pretrained with our noise-selection strategy exhibit significantly better tolerance, maintaining higher BLEU scores even as noise intensity increases. This performance stands in stark contrast to models pretrained without noise, which demonstrate a pronounced decline in BLEU scores under similar noise conditions.

\section{Conclusion}
\label{sec:conclusion}
In conclusion, this paper presents an innovative approach to text-in-image watermarking by harnessing the power of deep learning technologies. 
Our method utilizes Transformer-based models for text processing and Vision Transformers for analyzing image features, establishing a pioneering framework in the field. 
The deployment of deep learning techniques has enabled our model to exhibit enhanced adaptability, seamlessly adapting to the difference of images and the evolving landscape of digital security threats. Through extensive testing, our approach has been proven to offer superior robustness, achieving a level of imperceptibility that maintains the original image's quality.

\newpage
\bibliographystyle{IEEEtran}
\bibliography{IEEEabrv,references}
\end{document}